\documentclass[journal,article,submit,pdftex,moreauthors]{Definitions/mdpi} 

\firstpage{1} 
\pubvolume{1}
\issuenum{1}
\articlenumber{0}
\pubyear{2021}
\copyrightyear{2021}
\externaleditor{Academic Editor: Amelia Carolina Sparavigna}
\datereceived{21 November 2021} 
\dateaccepted{15 December 2021} 
\datepublished{} 
\hreflink{https://doi.org/} 

\usepackage{multirow}
\usepackage{amssymb}
\usepackage{amsmath}
\usepackage{mathdots}
\usepackage{float}
\usepackage{subcaption}
\usepackage{ragged2e}
\usepackage[algo2e,ruled,vlined]{algorithm2e}
\usepackage{xcolor}

\Title{Information Retrieval: Recent Advances and Beyond}

\TitleCitation{Information Retrieval: Recent Advances and Beyond}

\Author{
 Kailash Hambarde $^{1}$*, \mbox{and Hugo Proen\c{c}a $^{1,2}$}}

\AuthorNames{Kailash Hambarde and Hugo Proen\c{c}a}

\AuthorCitation{Hambarde, K.; Proen\c{c}a, H.}

\address{%
$^{1}$\quad University of Beira Interior, 
 Portugal; kailas.srt@gmail.com \\
$^{2}$\quad
 IT: Instituto de Telecomunica\c{c}\~oes, Portugal; hugomcp@ubi.pt}

\corres{Correspondence: kailas.srt@gmail.com}

\abstract{In this paper, we provide a detailed overview of the models used for information retrieval in the first and second stages of the typical processing chain. We discuss the current state-of-the-art models, including methods based on terms, semantic retrieval, and neural. Additionally, we delve into the key topics related to the learning process of these models. This way, this survey offers a comprehensive understanding of the field and is of interest for for researchers and practitioners entering/working in the information retrieval domain.}

\keyword{information retrieval; dense representation, document retrieval, relevance ranking} 

\begin{document}

\section{Introduction}
In the present day, Information Retrieval (IR) plays a vital role in human daily life through its implementation in a range of practical applications, including searching the web, question answering systems, personal assistants, chatbots, and digital libraries among others. The primary objective of IR is to locate and retrieve information that is relevant to a user's query. The main goal of IR is to identify and retrieve information that is related to a user's query. As multiple records may be relevant, the results are often ranked according to their relevance score to the user's query.

The traditional text retrieval systems mainly rely on the matching of terms between the query and the documents. However, term-based retrieval systems have several limitations such as polysemy, synonymy, and lexical gaps between the query and the documents \cite{croft2010search}. In recent years, advancements in computing power and the availability of large labeled datasets have greatly impacted the field of Natural Language Processing (NLP) by enabling researchers to use deep learning techniques for a variety of innovations. These techniques have been used to improve the traditional text retrieval systems and overcome the limitations of term-based retrieval systems.

However, these techniques require significant amounts of data and computing resources. As a result, researchers are continually developing more advanced deep learning algorithms to meet these demands and achieve better results in NLP tasks \cite{lin2021pretrained}. With the use of these advanced deep learning algorithms, the performance of IR systems has been greatly improved, leading to more accurate and efficient retrieval of information for the end-users. Some of the advancements in deep learning techniques that have been applied to IR include neural network architectures such as convolutional neural networks (CNNs) \cite{kim2014convolutional} and recurrent neural networks (RNNs) \cite{liu2016recurrent}, as well as transfer learning and pre-training techniques \cite{devlin2018bert}. These methods have been used to improve the representation of text data and to enhance the ability of IR systems to understand natural language queries.
Additionally, attention-based mechanisms such as the Transformer architecture \cite{vaswani2017attention} have been used to improve the ability of IR systems to attend to important parts of the query and documents for matching. Furthermore, the use of pre-trained language models, such as BERT \cite{devlin2018bert} and GPT-2 \cite{radford2019language}, have been shown to improve the performance of IR systems by providing a better understanding of the semantics and context of natural language queries and documents.

\begin{figure}[H]
    \centering
    \includegraphics[width=.8\linewidth,height=8cm]{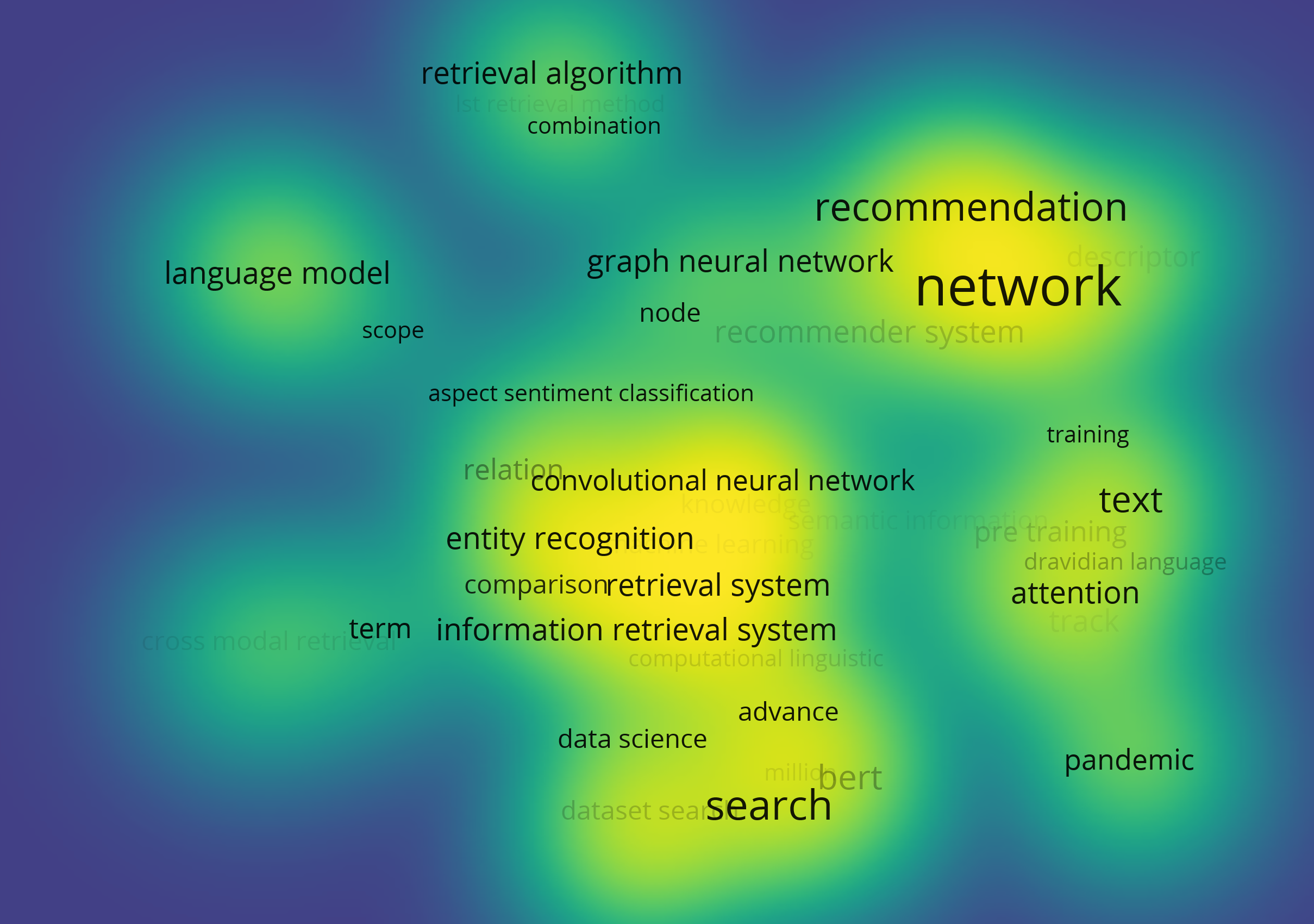}
    \caption{Term map of the information retrieval. Colors indicate the recent terms density, extracted from survery papers.}
    \label{fig:terms}
\end{figure}

The density term plot \ref{fig:terms}  provides a visual representation of the relative frequency of these keywords in the surveyed literature, which can be used to gain insights into the current research trends in the field.
Moreover, recent advancements in IR have also focused on incorporating external knowledge to improve the relevance of the retrieved information. For example, incorporating knowledge graph embeddings \cite{wang2017knowledge} into the IR process can help link the query and documents to the relevant entities and concepts, thus providing more accurate results. Furthermore, the use of multi-modal information retrieval, which combines text, image, and audio information has also been shown to improve the performance of IR systems \cite{wang2018learning}.

Overall, the advancements in deep learning techniques, large labeled datasets and high-computing power have significantly improved the performance of IR systems and made them more capable of handling the complexity of natural language queries. However, there is still room for further improvements and research in the field. In this paper, we aim to give a detailed overview of the models used for information retrieval in both the first and second stages of the process. We will discuss current models, including those based on terms, semantic retrieval and neural methods. We will also cover key topics related to the learning of these models. Our survey paper aims to provide a comprehensive understanding of the field and will be beneficial for researchers and practitioners in the area of information retrieval.

\section{Information Retrieval: Overview}
\begin{figure}[h]
    \centering
    \includegraphics[width=12cm, height=6cm]{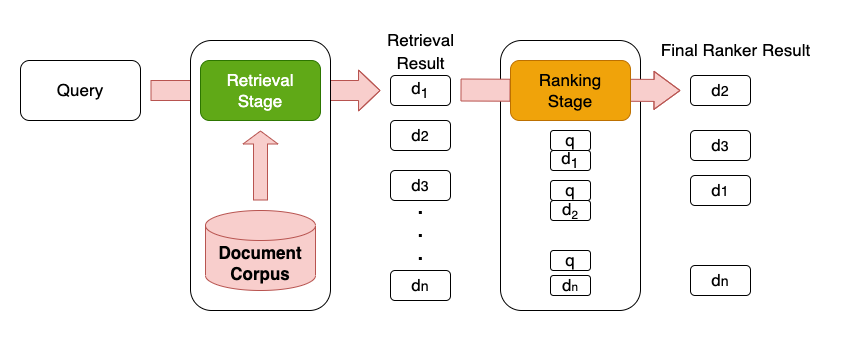}
    \caption{Overview of modern Information Retrieval system.}
    \label{fig:IRoverview}
\end{figure}

The basic objective of modern Information Retrieval (IR) is to provide the most relevant information to the end user in their query. This task can be divided into two parts: 1) retrieval; and 2) ranking, as illustrated in Fig. \ref{fig:IRoverview}. The first stage retrieves a set of initial documents that are likely to be relevant to the query, and the second stage re-ranks the rank documents based on their relevance score.

In terms of retrieval, the goal is to retrieve a set of relevant documents from the collection based on the user's query. This is done by using various algorithms and models such as the vector space model \cite{salton1975vector}, Boolean model, Latent Semantic Indexing (LSI) \cite{deerwester1990indexing}, Latent Dirichlet Allocation (LDA) \cite{blei2003latent} and recent techniques such as pre-trained models like BERT \cite{devlin2018bert}.

In the second stage, ranking, the main objective is to adjust the ranking of the initially retrieved documents based relevance score. The ranking process typically employs different models compared to those used in the retrieval stage, as the primary focus is on improving the effectiveness of the results rather than their efficiency.
Traditional models such as BM25 \cite{robertson1995okapi} are used as initial retrievers as they prioritize efficiency in recalling relevant documents from a massive document pool. Conventional ranking models encompass a range of techniques, such models include RankNet \cite{burges2005learning} and \cite{burges2006learning} for learning to rank and DRMM \cite{guo2016deep} and Duet \cite{mitra2017learning} for neural models. These models leverage various techniques, including reinforcement learning \cite{zhou2020rlirank}, contextual embeddings \cite{macavaney2019cedr} and attention mechanisms \cite{li2022learning}, to learn how to rank documents based on the user's query criteria.

\section{Conventional Term-based Retrieval}

Classical term-based retrieval, also known as Boolean retrieval, is a traditional approach to information retrieval that matches the terms in a query to the terms in a document \cite{manning2008introduction}. It is simple, fast, and easy to implement but has limitations such as the inability to handle synonyms, polysemy, and context \cite{baeza2011modern, ribeiro2011modern}. Despite these limitations, it is widely used in practical applications, especially in small-scale information retrieval systems \cite{manning2008introduction}. Other methods, such as the BIM proposed by \cite{robertson1976relevance} and the TDM proposed by \cite{van1977theoretical}, have been proposed to improve the accuracy and efficiency of retrieval systems. The TF-IDF weighting method is also an important concept in the field of IR, as proposed by \cite{salton1988term} and \cite{salton1991developments}. Recently, there has been a growing interest in using probabilistic and language modeling techniques to improve retrieval performance, such as the Query Likelihood model (QL) proposed by \cite{ponte2998language} and the divergence from randomness (DFR) approach introduced by \cite{amati2002probabilistic}.

\subsection{Pioneering Methods for Semantic Retrieval}

\subsubsection{Query Augmentation}
Early methods for information retrieval focused on using query expansion techniques to improve information retrieval systems. One approach was using global models to identify and utilize concepts relevant to a query \cite{qiu1993concept}. Another approach was using lexical-semantic relations to identify semantically related terms to the query \cite{voorhees1994query}. Additionally, using query context to improve retrieval performance was also studied \cite{bai2007using}. These methods demonstrate the potential of query expansion techniques, concepts and lexical-semantic relations to improve information retrieval systems.
Another important approach for query expansion is using local models, which incorporate user feedback into the retrieval process. A widely used local model is the Rocchio relevance feedback algorithm \cite{rocchio1971relevance}. It forms the foundation for many modern relevance feedback techniques.
\cite{zhai2001model} proposed incorporating user feedback into information retrieval systems using a divergence minimization model. \cite{cao2008selecting} presented a method for improving the effectiveness of pseudo-relevance feedback (PRF) in information retrieval systems. \cite{lv2009comparative} presented a study comparing various methods for incorporating PRF into information retrieval systems and \cite{zamani2016pseudo} proposed a novel approach for incorporating PRF into information retrieval systems using matrix factorization. These studies show the potential of using user feedback to improve retrieval performance.

\subsubsection{Document Augmentation}
Document expansion is a technique used in information retrieval (IR) systems to improve the performance of retrieval by expanding the representation of each document. The idea behind document expansion is that by including additional related terms in the representation of each document, the IR system will be able to better match the query with the relevant documents. \cite{kurland2004corpus} and \cite{liu2004cluster} focus on the relationship between corpus structure, language models, and ad-hoc information retrieval, with \cite{liu2004cluster} proposing a novel approach using clustering techniques. \cite{billerbeck2005document} compares the effectiveness of document expansion and query expansion methods in ad-hoc information retrieval systems. \cite{tao2006language} proposes a technique for expanding the representation of each document using related terms to improve retrieval performance. \cite{agirre2010document} expands documents using WordNet, a large lexical database of English. \cite{efron2012improving} focuses on improving the retrieval of short texts by expanding the representation of each document with semantically related terms. Lastly, \cite{sherman2017document} expands the representation of each document using external collections like WordNet. These papers show a progression in the field, with later papers building on the ideas presented in earlier papers and continuing to explore new methods for document expansion in information retrieval systems.

\subsubsection{Lexical Dependency Model}
The field of information retrieval has seen a significant amount of research focused on the use of lexical dependency models for document retrieval. One study that has contributed to this area is \cite{fagan1988experiments}, which compares the effectiveness of different automatic phrase indexing methods. Another important contribution to this field is \cite{salton1988term}, which presents a thorough analysis of various term weighting schemes and their impact on retrieval performance, and proposes a novel term dependency weighting scheme that accounts for the relationship between terms in a document.

Several other papers have built on the ideas presented in \cite{salton1988term} and \cite{fagan1988experiments}. For example, \cite{mitra1997analysis} proposes a new method for phrase-based retrieval using a vector space model (VSM) and term dependency weighting scheme. Similarly, \cite{song1999general} presents a new approach to information retrieval based on a general language model (LM) that incorporates term dependency weighting.
The use of probabilistic models for information retrieval has also been explored in the literature, with \cite{jones2000probabilistic} presenting a new probabilistic model that is based on the vector space model (VSM) and incorporates term dependency weighting. Another approach that has been proposed is the use of sentence trees to capture term dependencies, as presented in \cite{nallapati2002capturing}.
Recently, \cite{gao2004dependence} presents a new approach to information retrieval based on a language model (LM) that incorporates term dependency by proposing a new Dependence Language Model (DLM) that utilizes term dependencies to improve retrieval effectiveness. Furthermore, \cite{xu2010relevance} presents a new approach to relevance ranking in information retrieval that combines the use of kernels with the well-known BM25 retrieval function, and describes how the kernel-based approach can be used to model term dependencies similar to the schemes proposed in previous papers such as \cite{salton1988term} and \cite{mitra1997analysis}.

\subsubsection{Topic Model}
 Topic models have been widely used in the field of information retrieval as a way to improve the representation and indexing of text documents. One of the early contributions to this area was \cite{wong1985generalized}, which introduced the concept of Generalized Vector Space Model (GVSM) for information retrieval. This model extends the traditional vector space model (VSM) by allowing non-binary values in the term-document matrix.
Another important contribution to this field is Latent Semantic Analysis (LSA) for information retrieval, which was introduced in \cite{deerwester1990indexing}. LSA can be used to analyze and represent the semantic structure of a corpus of text documents for indexing and retrieval.
In recent years, various topic models have been proposed for information retrieval. For example, \cite{kurland2004corpus} describes how corpus statistics can be used to improve ad-hoc information retrieval systems. \cite{diaz2005regularizing} presents a new smoothing technique for ad-hoc retrieval based on regularizing the retrieval scores to account for the uncertainty in the language model. \cite{wei2006lda} introduces a new approach to ad-hoc retrieval that utilizes Latent Dirichlet Allocation (LDA) for both the indexing and the language modeling stages. \cite{yi2009comparative} compares the performance of various topic models for information retrieval, including LDA and PLSA. \cite{lu2011investigating} provides a comparison of the performance of PLSA and LDA on several benchmark datasets.
However, not all studies have found topic models to be effective for information retrieval. For example, \cite{atreya2011latent} critiques the performance of Latent Semantic Indexing (LSI) on TREC collections and argues that LSI is not effective for this type of data.

\subsubsection{Multilingual Retrieval Model}
Multilingual retrieval models have been an area of active research in the field of information retrieval. One of the key studies in this area is \cite{berger2017information}, which presents a new approach to information retrieval that views the retrieval task as a statistical translation problem and proposes a model to improve retrieval effectiveness.
Another important contribution to this field is \cite{karimzadehgan2010estimation}, which presents a new approach to estimating statistical translation models for ad-hoc information retrieval based on mutual information. \cite{gao2010clickthrough} describes a new approach to web search that utilizes click-through data to estimate translation models.
In addition to these studies, there have also been several papers that provide a theoretical analysis of the translation language model for information retrieval. For example, \cite{karimzadehgan2012axiomatic} presents an axiomatic analysis of the model, its properties and limitations. \cite{riezler2010query} describes a new approach to query expansion that utilizes monolingual statistical machine translation (SMT) to improve retrieval effectiveness. \cite{gao2012towards} describes a new approach to query expansion that utilizes concept-based translation models and search logs to improve retrieval effectiveness.

These papers demonstrate the potential of using statistical translation models to improve information retrieval systems.

\label{sec:back}
\unskip
\section{First Stage: Retrieval}
In this section, an extensive review of the literature pertaining to the first stage of information retrieval is conducted. The review is divided into three categories, namely: sparse techniques for semantic retrieval, state-of-the-art deep learning techniques for semantic retrieval, and hybrid techniques.

\subsection{Deep Learning Methods for Semantic Retrieval }
Neural-based approaches for semantic retrieval, such as those utilizing CNNs and RNNs have been proposed as a means to improve the effectiveness of information retrieval systems. These approaches involve representing text documents and queries in a continuous vector space, allowing for the utilization of neural network-based similarity measures, such as cosine similarity or dot-product, to rank documents according to their relevance to a given query \cite{shen2014latent, huang2013learning}. Furthermore, these neural methods can be used to learn complex non-linear representations of text data, which can be utilized to improve the performance of retrieval models \cite{kim2014convolutional, huang2013learning}.
Additionally, attention-based mechanisms such as the Transformer architecture \cite{vaswani2017attention} have been used to improve the ability of IR systems to attend to important parts of the query and documents for matching. Furthermore, the use of pre-trained language models, such as BERT \cite{devlin2018bert} and GPT-2 \cite{radford2019language}, have been shown to improve the performance of IR systems by providing a better understanding of the semantics and context of natural language queries and documents.

The use of neural methods in information retrieval has gained popularity in recent years, as evidenced by the increasing number of publications in this field \cite{wang2018neural, huang2013learning}.

\subsubsection{Discrete Retrieval Methods}

Sparse retrieval methods refer to techniques used in information retrieval that aim to reduce the dimensionality of the data being indexed and retrieve only the most relevant documents. These methods are based on the idea that the majority of the data in a collection is not relevant to a given query, and that it is unnecessary to retrieve all the documents when only a small subset is relevant.

One approach that has been proposed is the use of \textbf{term re-weighting methods}, which aim to assign higher weights to more relevant terms in the retrieval process.
One such method is "DeepTR" presented in \cite{zheng2015learning}, which learns to reweight terms based on their semantic similarity using distributed representations. Another approach is the integration of neural word embeddings into information retrieval systems, as described in \cite{zuccon2015integrating}. The authors propose a method for evaluating the effectiveness of different neural word embeddings in information retrieval and describe how these embeddings can be integrated into retrieval models.
Context-aware term weighting methods, such as "DeepCT" presented in \cite{dai2019context}, aim to improve the effectiveness of the first-stage retrieval process by assigning higher weights to more relevant terms based on their context. The authors propose a deep learning-based method that learns to estimate the importance of terms in a sentence or passage based on their context.
"TDV" presented in \cite{frej2020learning} is another approach that focuses on learning term discrimination in information retrieval. The authors propose a deep learning-based method that learns to distinguish relevant terms from irrelevant terms based on their context.
Building on the concept of "DeepCT", "HDCT" proposed in \cite{dai2020context} considers the term importance in multiple levels of granularity, from document-level to sentence-level, to improve the effectiveness of the retrieval process. The efficiency of "DeepCT" approach is evaluated and compared with traditional term weighting methods in terms of computational cost and retrieval performance in \cite{mackenzie2020efficiency}.

Finally, \cite{lin2021few} presents a conceptual framework for analyzing and comparing different information retrieval techniques, particularly neural-based methods. They introduce "DeepImpact" and "COIL" (Continuous Optimization-based Information Retrieval) as two different dimensions of the framework, and propose "uniCOIL" (Unified Continuous Optimization-based Information Retrieval) as a unified approach that combines the strengths of both dimensions.

\textbf{Expansion} Information retrieval systems have been constantly evolving in order to improve their effectiveness. One approach that has been proposed is expanding a given document or query by predicting or generating relevant queries, and then using those queries to retrieve additional information.
\cite{nogueira2019document} proposed a model called "Doc2Query" which uses a neural network to predict the relevant queries given a document, and then retrieves additional documents using those queries. They showed that this approach is able to improve the effectiveness of document retrieval.
\cite{nogueira2019doc2query} further improved upon their previous work by incorporating additional information, such as the title of the document and the clicked snippets, into the Doc2Query model. They also proposed a new evaluation metric for the task, which they called "Top-k-TTTTT" (where TTTTT stands for "title, terms, terms, terms, terms"). They showed that the new model, called "DocTTTTTQuery", outperforms the previous model on the benchmark dataset.
\cite{mao2020generation} proposed a model called "GAR" (Generation-augmented Retrieval) which generates new queries based on the given question and then retrieves relevant documents using those queries. They showed that this approach is able to improve the effectiveness of open-domain question answering.
\cite{yan2021unified} proposed a model called "UED" (Unified Expansion and Ranking model) that is trained on a large corpus of text using a combination of supervised and unsupervised learning. They showed that this model is able to effectively rank and expand passages. \cite{yan2021unified,mao2020generation,nogueira2019doc2query,nogueira2019document} proposed new methods for expanding a given document or query by predicting or generating relevant queries, and then using those queries to retrieve additional information. They show that these methods can improve the effectiveness of document retrieval, open-domain question answering, and passage ranking. With each new approach, the field of information retrieval continues to evolve and improve.

\textbf{Expansion \'+' Term Re-weighting}
Recent research in information retrieval has focused on using a combination of expansion and term re-weighting techniques to enhance the performance of retrieval systems, such as EPIC proposed by \cite{macavaney2020expansion} which uses a neural network to predict the importance of terms in a document and then uses those terms to retrieve additional information, SparTerm proposed by \cite{bai2020sparterm} which uses a neural network to learn a sparse term-based representation of a document, SPLADE and SPLADE v2 proposed by \cite{formal2021splade} which use neural networks to learn a sparse lexical representation of a document and expand the query, DeepImpact proposed by \cite{mallia2021learning} which uses a neural network to learn the impact of passages on the retrieval process, TILDEv2 proposed by \cite{zhuang2021fast} which uses a neural network to learn term-based representations of passages and expand the query, and SpaDE proposed by \cite{choi2022spade} which uses a neural network to learn a sparse representation of a document and encode the document using two encoders. 

\textbf{Sparse Representation Learning} 
Another approach that has been proposed to improve the effectiveness and efficiency of information retrieval systems is the use of sparse representations.
\cite{salakhutdinov2009semantic} proposed a method called semantic hashes, which use a neural network to map documents to compact binary codes that can be used for efficient approximate nearest neighbor search. They showed that this approach is able to achieve state-of-the-art performance.
\cite{zamani2018neural} proposed a model called SNRM which uses a neural network to learn a sparse representation of a document and showed that it is able to improve the effectiveness of information retrieval.
\cite{jang2021ultra} proposed a model called UHD-BERT which uses a neural network to learn ultra-high-dimensional sparse representations of a document and showed that it is able to improve the effectiveness of full-ranking.
\cite{yamada2021efficient} proposed a model called BPR which uses a neural network to learn a semantic hash for each passage and showed that it is able to improve the efficiency of open-domain question answering.
\cite{lassance2021composite} proposed a model called CCSA which uses a neural network to learn a composite code sparse representation of a document and showed that it is able to improve the effectiveness of first-stage retrieval. All these papers propose new methods that use sparse representations to improve the effectiveness and efficiency of information retrieval systems.

\subsection{Dense Retrieval Methods}
In recent years, word embeddings have become a popular method for representing documents and queries in information retrieval systems. These embeddings are dense, continuous representations of words that capture their semantic meaning.
\textbf{Word-Embedding-based}
\cite{clinchant2013aggregating} uses word embeddings and a Fisher Vector aggregation technique to represent documents and queries, \cite{vulic2015monolingual} uses bilingual word embeddings for cross-lingual information retrieval, \cite{kenter2015short} uses word embeddings to represent short texts, \cite{mitra2016dual} uses two different embedding spaces to represent documents and queries, and \cite{henderson2017efficient} uses word embeddings to represent context and potential responses for natural language generation.  \cite{gillick2018end} and \cite{karpukhin2020dense} both use neural networks to map documents and queries to a continuous space and apply a similarity measure to rank the documents, while \cite{seo2018phrase} uses a neural network to extract important phrases from documents and indexes them to improve efficiency in retrieval. 
\cite{lewis2020retrieval}, suggests using a retrieval component to gather relevant information and a generation component to create a response, while \cite{zhan2020repbert} proposes using contextualized text embeddings to improve first-stage retrieval. \cite{wrzalik2020cort} suggests using multiple transformer-based models to improve document retrieval.
 One approach that has been gaining popularity is the use of pre-trained transformer-based models for encoding questions and documents. \cite{nie2020dc} presents a method for document retrieval by decoupling the encoding of questions and documents using a pre-trained transformer-based model called "DC-BERT". The authors show that this approach is able to improve the effectiveness of document retrieval. Similarly, \cite{yang2020neural} presents a method for question answering that uses a neural retrieval component and a cross-attention component to generate a response based on the retrieved passages. They also propose a data augmentation method to improve the performance of the model.

Another approach that has been explored is the use of approximate nearest neighbor search and negative contrastive learning for dense text retrieval. \cite{xiong2020approximate} presents a method called ANCE which uses a neural network to encode documents and queries, and then applies approximate nearest neighbor search and negative contrastive learning to rank the documents. 
 \cite{zhan2020learning} proposes a method for training dense retrieval models effectively and efficiently, using a combination of hard and soft negative sampling, and optimizing the balance between the retrieval and generation components of the model. \cite{shan2020bison} presents a method for web search using a global weighted self-attention network, which improves the effectiveness of web search. \cite{qu2020rocketqa} presents a method for open-domain question answering using dense passage retrieval and an optimized training approach, which improves the effectiveness of open-domain question answering.

The papers \cite{zhan2020learning}, \cite{shan2020bison}, \cite{qu2020rocketqa}, \cite{lee2020learning}, \cite{hofstatter2021efficiently}, \cite{zhan2021optimizing}, \cite{yu2021few}, and \cite{li2021more} all propose different methods for improving the performance of dense retrieval models. \cite{zhan2020learning} focuses on effective and efficient training using a combination of hard and soft negative sampling, \cite{shan2020bison} uses a global weighted self-attention network for web search, \cite{qu2020rocketqa} uses dense passage retrieval and an optimized training approach for open-domain question answering, \cite{lee2020learning} uses dense representations of phrases to improve performance, \cite{hofstatter2021efficiently} uses balanced topic aware sampling to improve the quality of the training data, \cite{zhan2021optimizing} uses hard negatives to improve the model's ability to handle difficult cases, \cite{yu2021few} is focused on training dense retrieval models for conversational scenarios with a limited number of examples, and \cite{li2021more} propose more robust to variations in the input data.

Recently, researchers have been exploring new methods to improve the performance of dense retrieval models, which are used to retrieve relevant information from a large collection of documents. In \cite{ren2021pair}, the authors propose a new method for dense passage retrieval by using a passage-centric similarity relation, which focuses on the relationship between passages rather than individual words.
Building on this, \cite{khattab2021relevance} proposed a new method for training ColBERT, a pre-trained transformer model, for OpenQA by using relevance-guided supervision. This approach focuses on training the model to learn the relevance of passages to a given query, which is important for OpenQA tasks.
In \cite{singh2021end}, the authors propose a new method for training a multi-document reader and retriever for open-domain question answering by using end-to-end training. This approach focuses on training the entire system, including the reader and retriever components, in an end-to-end manner to improve performance.
The paper \cite{yu2021improving} propose a new method for improving query representations for dense retrieval by using pseudo relevance feedback, which is a technique to extract relevant information from a set of retrieved documents and use it to improve the retrieval performance.

Following this, the paper \cite{wang2021pseudo} propose a new method for training dense retrieval models by using pseudo-relevance feedback and multiple representations, which allows the model to learn more robust representations of queries.
To further improve performance, \cite{cai2021discriminative} proposed a new method for training a discriminative semantic ranker for question retrieval. This approach focuses on training the model to differentiate between relevant and non-relevant documents, which is important for accurate retrieval.

Finally, in \cite{wu2021representation}, the authors propose a new method for improving open-domain passage retrieval by using representation decoupling, which separates the encoding of passages and queries to improve the model's ability to understand the relationships between them. In \cite{ren2021rocketqav2}, the authors introduce a new approach for training a dense passage retrieval and re-ranking model.
\cite{lindgren2021efficient} presents a method for more efficiently training retrieval models through the use of negative caching.
\cite{lu2021multi} proposes a technique for training neural passage retrieval models by utilizing multi-stage training and improved negative contrast.All these methods were proposed to improve the effectiveness of dense retrieval models, and they all showed promising results in their respective evaluation.

\textbf{Knowledge Distillation} is a technique for transferring knowledge from a pre-trained, larger model to a smaller model. This can be done to improve the performance of the smaller model on a specific task, or to make the model more efficient in terms of computational resources. Researchers propose methods for transferring knowledge from a pre-trained, larger model to a smaller model in order to improve performance on specific tasks such as document ranking \cite{lin2020distilling}, chat-bot systems \cite{vakili2020distilling}, question answering \cite{izacard2020distilling} and large-scale retrieval tasks \cite{lu2020twinbert}. These methods include the use of a pre-trained BERT model \cite{lin2020distilling}; \cite{lu2020twinbert}, transferring knowledge across different model architectures and using a Margin-MSE loss function \cite{hofstatter2020improving}. The papers also explore the relationship between pre-trained models and the effect of distilling knowledge from one model to another \cite{yang2020retriever} \cite{choi2021improving}.

\textbf{Multi-vector Representation}

In \cite{feldman2019multi}, the authors propose a method called MUPPET which uses multiple hops to retrieve paragraphs that are relevant to a given question.
In \cite{luan2021sparse}, the authors propose a method called ME-BERT, which uses a combination of sparse, dense, and attentional representations to improve text retrieval tasks.
In \cite{khattab2020colbert}, the authors propose a method called ColBERT which uses contextualized late interaction to improve passage search.
In \cite{gao2021coil}, the authors propose a method called COIL which uses contextualized inverted lists to improve exact lexical match in information retrieval.
In \cite{tang2021improving}, the authors propose a method that generates pseudo query embeddings to improve the representation of documents for dense retrieval.
In \cite{lee2021phrase}, the authors propose a method called DensePhrases, which uses phrase retrieval to learn passage retrieval.
In\cite{tonellotto2021query}, the authors propose a method for pruning query embeddings to improve dense retrieval performance.
In \cite{zhang2022multi}, the authors propose a method for learning multi-view document representations to improve open-domain dense retrieval.
In \cite{li2022learning}, the authors propose a method for learning diverse document representations by using deep query interactions.
In \cite{du2022topic}, the authors propose a method for using topic-grained text representations to improve document retrieval.

\textbf{Accelerate Interaction-based Models}
Accelerating interaction-based models for information retrieval tasks is a crucial area of research in natural language processing. There have been several recent studies that propose various methods for improving the efficiency and accuracy of open-domain question answering and document retrieval.
In \cite{mitra2019incorporating, mitra2020conformer}, the authors propose methods that incorporate the query term independence assumption to improve retrieval and ranking using deep neural networks.
\cite{ji2019efficient} proposes a method for efficient interaction-based neural ranking using locality sensitive hashing.
\cite{humeau2019poly} proposes a method called Poly-encoders, which uses a combination of architectures and pre-training strategies for fast and accurate multi-sentence scoring.
\cite{gao2020modularized} proposes a method called MORES, which uses a modularized transformer-based ranking framework to improve document retrieval.
\cite{macavaney2020efficient} proposes a method called PreTTR, which uses precomputing term representations to improve document re-ranking for transformers.
\cite{cao2020deformer} proposes a method called DeFormer, which decomposes pre-trained transformers for faster question answering.
\cite{zhao2020sparta} proposes a method called SPARTA, which uses sparse transformer matching retrieval for efficient open-domain question answering.

\textbf{Pre-training}
In recent years, there has been a growing interest in pre-training methods for dense passage retrieval, which aims to improve the performance of retrieval systems by leveraging large amounts of unannotated data. In this survey, we will discuss several methods proposed in the literature that utilize pre-training to improve dense passage retrieval.
One line of research focuses on using latent retrieval for pre-training. For example, in \cite{lee2019latent}, the authors propose a method called ORQA, which uses latent retrieval for weakly supervised open domain question answering. Similarly, in \cite{guu2020retrieval}, the authors propose a method called REALM, which uses retrieval-augmented language model pre-training to improve performance.
Another line of research focuses on using pre-training tasks to improve embedding-based large-scale retrieval. For example, in \cite{chang2020pre}, the authors propose a method called BFS+WLP+MLM, which uses pre-training tasks to improve embedding-based large-scale retrieval.
Additionally, there are methods that use dense representation fine-tuning for language models. For example, in \cite{gao2021your}, the authors propose a method called Condenser, which uses dense representation fine-tuning for language models. Similarly, in \cite{gao2021unsupervised}, the authors propose a method called coCondenser, which uses unsupervised corpus aware language model pre-training for dense passage retrieval.
Another line of research focuses on using a weak decoder to pre-train a strong siamese encoder. For example, in \cite{lu2021less}, the authors propose a method called SEED-Encoder, which uses a weak decoder to pre-train a strong siamese encoder.
Additionally, there are methods that use hyperlink information for pre-training in ad-hoc retrieval. For example, in \cite{ma2021pre}, the authors propose a method called HARP, which uses hyperlink information for pre-training in ad-hoc retrieval.
Another line of research focus on using contextual information to improve dense passage retrieval. For example, in \cite{wu2022contextual}, the authors propose a method called ConTextual Mask Auto-Encoder (CMAE), which uses contextual information to improve dense passage retrieval.
Additionally, there are methods that use masked auto-encoder for pre-training retrieval-oriented transformers. For example, in \cite{liu2022retromae}, the authors propose a method called RetroMAE, which uses masked auto-encoder for pre-training retrieval-oriented transformers.
Another line of research focus on using representation bottleneck for pre-training dense passage retrieval models. For example, in \cite{wang2022simlm}, the authors propose a method called SimLM, which uses representation bottleneck for pre-training dense passage retrieval models.
Lastly, there are methods that use lexicon-bottlenecked pre-training for large-scale retrieval. For example, in \cite{shen2022lexmae}, the authors propose a method called LexMAE, which uses lexicon-bottlenecked pre-training for large-scale retrieval. The last method is a method that uses a contrastive pre-training approach to learn discriminative autoencoder for dense retrieval. For example in \cite{ma2022contrastive}.
Overall, these pre-training methods demonstrate promising results in improving dense passage retrieval performance. However, more research is needed to fully understand

\textbf{Zero-shot/Few-shot Learning}
Recent research in the field of zero-shot information retrieval has been focused on finding ways to improve the performance and generalizability of retrieval models. One common approach is the use of query generation, as seen in \cite{liang2020embedding} and QGen \cite{ma2020zero}. These methods aim to improve zero-shot retrieval by embedding queries into a shared space, or by using synthetic question generation to improve passage retrieval.
Another approach is the use of synthetic pre-training, as suggested by \cite{reddy2021towards}, to improve the robustness of neural retrieval models. Additionally, \cite{thakur2021beir} presents a benchmark for evaluating the performance of zero-shot retrieval models.
Another approach is the use of momentum adversarial domain-invariant representations as proposed by \cite{xin2021zero} for zero-shot dense retrieval. \cite{ni2021large} proposes DTR, which uses large dual encoders for generalizable retrieval. \cite{chen2022out} suggests utilizing out-of-domain semantics to improve zero-shot hybrid retrieval models. \cite{bonifacio2022inpars} proposes InPars, a method that uses large language models for data augmentation in information retrieval.

Moreover, contrastive learning has been proposed as a means of performing unsupervised dense information retrieval with Contriever method by \cite{izacard2021towards}. \cite{wang2021gpl} proposes GPL method that uses generative pseudo-labeling for unsupervised domain adaptation of dense retrieval. \cite{ram2021learning} proposes a method called Spider, which enables unsupervised passage retrieval. \cite{neelakantan2022text} proposes a method that utilizes contrastive pre-training to learn embeddings for text and code. \cite{zhan2022disentangled} proposes a method that disentangles the modeling of domain and relevance for adaptable dense retrieval. Lastly, \cite{dai2022promptagator} proposes a method called Promptagator, which uses few-shot dense retrieval from 8 examples.

Overall, recent research in zero-shot information retrieval has been focused on finding ways to improve the performance and generalizability of retrieval models by utilizing various techniques such as query generation, synthetic pre-training, large language models, and contrastive learning. These methods have been applied to tasks such as passage retrieval, unsupervised domain adaptation, and dense retrieval.

\textbf{Probing Analysis}
Several studies have been conducted to address the limitations of dense low-dimensional retrieval, particularly when dealing with large index sizes. These studies propose various methods such as redundancy elimination, benchmarking, incorporating salient phrase information, entities-centric questions, and interpreting dense retrieval as a mixture of topics as ways to improve the performance and interpretability of dense retrieval models. For example, \cite{reimers2020curse} delves into the challenges that arise from utilizing dense low-dimensional retrieval for large index sizes and proposes solutions to mitigate these limitations. Similarly, \cite{ma2021simple} proposes a method for unsupervised redundancy elimination to compress dense vectors for passage retrieval as a means of enhancing retrieval performance. \cite{thakur2021beir} proposed a benchmark to evaluate the performance of zero-shot information retrieval models, while \cite{chen2021salient} explores the potential of incorporating salient phrase information in dense retrieval to imitate the performance of sparse retrieval. \cite{sciavolino2021simple} proposed a benchmark of simple entities-centric questions to evaluate and challenge the performance of dense retrievers, and \cite{zhan2021interpreting} proposed a new approach to improve the effectiveness and interpretability of dense retrieval by interpreting it as a mixture of topics.

\subsection{Hybrid Retrieval Methods}

\textbf{Word-Embedding-based}
\cite{vulic2015monolingual} proposed a method for linearly combining monolingual and cross-lingual word embeddings to improve retrieval performance.
\cite{ganguly2015word} proposed a GLM (Generalized Language Model) which utilizes word embeddings to improve the retrieval performance.
\cite{roy2016representing} proposed a method for combining word embeddings using set operations to improve retrieval performance.
\cite{mitra2016dual} proposed a method for combining word embeddings into a dual embedding space model (DESM) to improve retrieval performance.
\cite{boytsov2016off} proposed a method for replacing term-based retrieval with k-NN search and incorporating translation models and BM25 to improve retrieval performance.
\cite{dos2015learning} proposed a method for combining Bag-of-Words (BOW) and CNN (Convolutional Neural Network) to improve question answering performance.
\cite{seo2019real} proposed a method called DenSPI (Dense-Sparse Phrase Index) to improve question answering performance in real-time.
\cite{lee2019contextualized} proposed a method called SPARC (Sparse, Contextualized Representations) to improve question answering performance in real-time.
\cite{wrzalik2020cort} proposed a method called CoRT (Complementary Rankings from Transformers) which combines transformer-based models with BM25 to improve retrieval performance.
\cite{luan2021sparse} proposed a method called ME-Hybrid, which combines sparse and dense representations with attentional mechanisms to improve retrieval performance.
\cite{gao2021complement} proposed a method called CLEAR (Complement Lexical Retrieval Model) which combines lexical and semantic residual embeddings to improve retrieval performance.
\cite{kuzi2020leveraging} proposed a hybrid approach that combines semantic and lexical matching to improve recall of document retrieval systems.
\cite{lin2021few} proposed a conceptual framework for information retrieval techniques called uniCOIL (unified Conceptual framework for Information Retrieval)
\cite{chen2021contextualized} proposed a method called CORW (Contextualized Offline Relevance Weighting) which improves the efficiency and effectiveness of neural retrieval by utilizing context and relevance weighting.
Overall, these papers show that combining different methods, representations and architectures can lead to improved performance of question answering and text retrieval systems. They propose different methods such as hybrid representations, dense-sparse phrase index, sparse and dense representations, attentional mechanisms, lexical and semantic embeddings, semantic and lexical matching and offline relevance weighting.
\cite{arabzadeh2021predicting} proposed a method for predicting efficiency and effectiveness trade-offs for dense vs. sparse retrieval strategies.
\cite{leonhardt2021fast} proposed a method called Fast Forward Indexes which aims to improve the efficiency of document ranking by using a forward index that stores the positions of terms within documents.
\cite{lin2021densifying} proposed a method called Representational Slicing which densifies sparse representations for passage retrieval.
\cite{shen2022unifier}) proposed a method called UniﬁeR which aims to improve the efficiency and effectiveness of large-scale retrieval by unifying different retrieval strategies.

Overall, these papers show that different methods can be used to improve the efficiency and effectiveness of text retrieval systems. They propose different methods such as contextualized offline relevance weighting, predicting efficiency/effectiveness trade-offs, fast forward indexes, densifying sparse representations, and unified retrieval for large-scale retrieval.

\section{Second Stage - Ranker}
In recent years, there has been a growing interest in the use of neural networks for IR ranking. These models, also known as deep ranking models, have been shown to be highly effective in learning the relevance of documents from labeled training data. One of the key advantages of using deep learning for IR ranking is the ability to learn complex representations of the documents and queries, which can capture subtle semantic relationships and handle large amounts of data \cite{mitra2017learning}.
The Vector Space Model (VSM) \cite{salton1975vector} is a widely used algorithm in information retrieval (IR) that represents documents and queries as vectors in a high-dimensional space and ranks documents based on their similarity to the query. The basic idea is that the smaller the angle between the vectors, the more similar the query and the document are. One of the first papers to propose the use of VSM for IR was \cite{salton1988term}, which proposed a probabilistic model for IR and showed that it outperforms the traditional Boolean model. Recently, there have been a number of papers that have proposed using VSM in combination with other IR algorithms to improve performance, such as \cite{desai2022comparative, guo2016deep, wei2006lda, zhang2021graph, ma2021prop, khodabakhsh2022qualitative, agarwal2019general, ai2019learning, boualili2020markedbert, capannini2016quality}.

Ad-hoc retrieval, where a user's query is used to search a collection of documents for relevance, is a fundamental task in the field of information retrieval \cite{baeza1999modern,mitra2017neural}. The task involves using a ranking algorithm to create an ordered list of documents from a corpus, where the top-ranked documents are deemed the most pertinent to the user's query. However, the search request in ad-hoc retrieval is typically brief and may come from a user with unclear intent, creating a vocabulary mismatch between the user's query and the terms present in the collection, hindering the re-ranker's ability to match relevant documents \cite{furnas1987vocabulary,zhao2010term}.

Recent research has focused on addressing these challenges by proposing novel approaches to ad-hoc retrieval. For example, \cite{guo2016deep} presents the deep relevance matching model (DRMM) which addresses the challenges of relevance matching through a joint deep architecture at the query term level. Other papers propose incorporating topic models like LDA \cite{wei2006lda} and graph neural networks \cite{zhang2021graph} to leverage document-level word relationships for ad-hoc retrieval. Additionally, pre-training methods such as PROP \cite{ma2021prop} and counterfactual inference techniques \cite{agarwal2019general} have been proposed to improve the performance of pre-trained language models in IR tasks. Furthermore, \cite{boualili2020markedbert} present a new approach to improve the performance of pre-trained language models in IR tasks by using established IR cues such as exact term-matching.

Overall, the field of information retrieval is constantly evolving, with researchers proposing new and innovative approaches to improve the performance of ad-hoc retrieval and other IR tasks. The above-mentioned papers highlights the various approaches that have been proposed in recent years to address the challenges of ad-hoc retrieval and improve the performance of IR algorithms.
\paragraph{}

IR researchers have proposed a variety of approaches to optimize and improve performance, with a particular focus on learning-to-rank (LTR) systems. One such approach is the use of cascaded ranking models, as proposed by in \cite{chen2017efficient}. The authors propose integrating feature costs into multi-stage LTR systems, which they demonstrate can increase both efficiency and effectiveness. Another approach is the use of weak supervision, as proposed by in \cite{agarwal2019general}. The authors present a method for training a neural ranking model using weak supervision, where labels are obtained automatically without human annotators or external resources.

\cite{fan2018modeling} presents a novel approach for assessing relevance between a query and a document in ad-hoc retrieval, which addresses the challenge of diverse relevance patterns. The proposed method is a data-driven approach called Hierarchical Neural matching model which consists of two stacked components: a local matching layer and a global decision layer. \cite{gao2020earl} propose a method for making transformer-based ranking models more efficient by modularizing them into separate modules for text representation and interaction. This makes the ranking process faster using offline pre-computed representations and lightweight online interactions. Additionally, this modular design makes the model easier to interpret and better understand the ranking process in transformer-based models.

Several studies have been conducted to make transformer-based models more efficient and effective. One such study, by \cite{gao2020modularized}, proposed a modularized design for transformer-based ranking models to improve efficiency by using offline pre-computed representations and lightweight online interactions. This design also improved interpretability by providing insight into the ranking process. Another study, by \cite{gao2021rethink}, presented a method for improving text retrieval performance using pre-trained deep language models (LMs) through a technique called Localized Contrastive Estimation (LCE). The LCE approach aims to address the issue that popular rerankers are not able to fully exploit improved retrieval results when pre-trained deep LMs are used to improve search index. The experiments showed that this approach significantly improves text retrieval performance. \cite{hofstatter2019tu,hofstatter2020improving,hofstatter2020interpretable} presented a balanced neural re-ranking approach for text retrieval called TK (Transformer-Kernel) model, which uses a small number of transformer layers to contextualize query and document word embeddings, and a document-length enhanced kernel-pooling to score individual term interactions.
Furthermore, \cite{hofstatter2020interpretable} presented TK (Transformer-Kernel), a neural re-ranking model for ad-hoc search that uses an efficient contextualization mechanism. The model's design aims to achieve an optimal balance between effectiveness and efficiency. \cite{jiang2020long} proposed the Query-Directed Sparse attention method for document ranking tasks in transformer models, called QDS-Transformer, which aims to improve efficiency while still maintaining important properties for ranking such as local contextualization, hierarchical representation, and query-oriented proximity matching.
\paragraph{}
\paragraph{}
Researchers in the field of information retrieval (IR) have proposed a variety of approaches to optimize and improve the performance of IR systems, with a particular focus on learning-to-rank (LTR) systems. One approach is the use of cascaded ranking models, as proposed by in \cite{chen2017efficient}. The authors propose integrating feature costs into multi-stage LTR systems, which they demonstrate can increase both efficiency and effectiveness. Another approach is the use of weak supervision, as proposed by in \cite{agarwal2019general}. The authors present a method for training a neural ranking model using weak supervision, where labels are obtained automatically without human annotators or external resources.
\cite{joachims2017unbiased} presented a counterfactual inference framework for unbiased LTR using implicit feedback data. They highlighted the issue of bias in using this type of data, specifically position bias in search rankings, and its negative impact on LTR methods. To overcome this problem, the authors proposed a Propensity-Weighted Ranking SVM which uses click models as the propensity estimator.
 \cite{lin2019impact} discussed the issue of repeatability in document ranking experiments using the open-source Lucene search engine. They pointed out that score ties during retrieval, which occur when multiple documents have the same ranking score, can lead to non-deterministic rankings due to multi-threaded indexing. They suggested using external collection document ids as a solution to ensure repeatability.
in \cite{lin2020distilling} presented an approach to improve the ranking performance of the late-interaction ColBERT model by applying knowledge distillation. The authors used distillation to transfer knowledge from the ColBERT model into a simpler dot product, which improves query latency and reduces storage requirements, while still maintaining good effectiveness.
in \cite{lovon2021studying} studied the cross-domain transferability of deep neural ranking models in IR and found that they "catastrophically forget" old knowledge when learning new knowledge from different domains, leading to decreased performance. The study suggests that using a lifelong learning strategy with a cross-domain regularizer can mitigate this problem and provides insights on how domain characteristics impact the issue of catastrophic forgetting.
 in \cite{macavaney2019cedr} investigated the use of contextualized language models (ELMo and BERT) for ad-hoc document ranking in information retrieval. They proposed a new joint approach, CEDR, which incorporates BERT's classification vector into existing neural models and showed that it outperforms current baselines. The paper also addressed practical challenges like maximum input length and runtime performance.
 in \cite{macavaney2019content} suggests using weak supervision sources to train neural ranking models. These sources provide pseudo query-document pairs that already exhibit relevance, such as newswire headline-content pairs and encyclopedic heading-paragraph pairs. The authors also proposed filters to eliminate irrelevant training samples. They showed that these weak supervision sources are effective and that filtering can further improve performance. This approach allows for more efficient and cost-effective training of neural ranking models.
in \cite{macavaney2020efficient} proposed PreTTR (Precomputing Transformer Term Representations) which precomputes part of the document term representations at indexing time and merges them with query representation at query time for faster ranking score computation. Additionally, the authors propose a compression technique to reduce storage requirements
\paragraph{}

Another approach is using curriculum learning to train neural answer ranking models \cite{macavaney2020training}. The idea is to learn to identify easy correct answers before incorporating more complex logic. Heuristics are proposed to estimate the difficulty of a sample and used to build a training curriculum that gradually shifts to weighting all samples equally. This approach has been shown to improve the performance of BERT and ConvKNRM models.
Standard neural rerankers have also been proposed to improve the efficiency of IR-based QA systems \cite{matsubara2020reranking}. These rerankers can reduce the number of sentence candidates for the Transformer models without affecting accuracy, improving efficiency by up to 4 times. This allows for the use of powerful Transformer models in real-world applications while maintaining accuracy.
Additionally, researchers have proposed new models for document relevance ranking, such as the Deep Relevance Matching Model (DRMM) \cite{mcdonald2018deep}, which have context-sensitive encodings throughout, inspired by another existing model, PACRR. These encodings are used in multiple ways to match query and document inputs. Furthermore, incorporating query term independence into neural IR models has also been proposed \cite{mitra2019incorporating} as a way to make them more efficient for retrieval from large collections. This approach has been shown to result in little loss in result quality for Duet and CKNRM and a small degradation in the case of BERT.

\cite{mokrii2021systematic} evaluates transferability of BERT-based neural ranking models across 5 datasets and finds that training on pseudo-labels can produce a competitive or better model than transfer learning. However, it also highlights the need to improve stability and effectiveness of few-shot training as it can degrade performance of pretrained model.

\cite{wu2020leveraging} proposes a new approach to document ranking called the context-aware Passage-level Cumulative Gain (PCG) which aggregates relevance scores of passages and accounts for context information. This work aims to improve ranking performance and provide more explainability for document ranking.

\section{Datasets}

Table \ref{Sec:dataset} presents a summary of the state-of-the-art datasets used in the survey paper. The table includes datasets for various NLP tasks such as Passage-Retrieval \cite{nguyen2016ms}, Bio-Medical Information Retrieval \cite{wang2020cord,tsatsaronis2015overview,boteva2016}, Question Answering \cite{kwiatkowski2019natural, yang2018hotpotqa,maia201818,khot2020qasc,feng2015applying}, Tweet\cite{suarez2018data}, News Article \cite{soboroff2018trec}\cite{voorhees2005trec}, Argument Retrieval \cite{wachsmuth2018retrieval,ajjour2020overview}, Community QA \cite{hofstatter2020improving}, Entities \cite{hasibi2017dbpedia}, Fact Checking  \cite{thorne2018fever,diggelmann2020climate,wadden2020fact} and Search Query \cite{qin2013introducing}. The table lists the task, dataset name, domain, and corpus size for each dataset. The datasets are organized into sections based on the task they are used for and the table provides a brief description of each dataset, including its source, as mentioned in the references, and the size of the corpus. This table can be used as a reference for researchers to select appropriate datasets for their NLP tasks and to cite the datasets used in their research.

\nointerlineskip
\setcounter{table}{1}
\begin{table}[H]
\scriptsize
\setlength{\tabcolsep}{7.8mm}
\caption{SOTA Datasets in IR. For each dataset, we provide the corresponding task, domain and corpus.}
\label{tab:obj3}
\setlength{\tabcolsep}{4mm}
\begin{tabular}{llcc}
\toprule
\textbf{Task} & \textbf{Data-set} & \textbf{Domain} & \textbf{Corpus}\\ 
\midrule
Passage-Retrieval & MS MARCO \cite{nguyen2016ms} & Real Queries & 8.84M\\ 
\midrule
Bio-Medical& CORD-19 \cite{wang2020cord} & Scholarly Articles & 171K\\
Information& BioASQ \cite{tsatsaronis2015overview} &Human-annotated QA &14.91M\\
Retriveval& NFCorpus \cite{boteva2016} & Medical Information & 3.6K\\
\midrule

& Natural Questions \cite{kwiatkowski2019natural} & Real Users QA & 2.68M\\
Question & HotpotQA \cite{yang2018hotpotqa} & Question Answering QA & 5.23M\\
Answering & FiQA-2018	 \cite{maia201818} & Financial Opinion QA & 57K\\
& QASC \cite{khot2020qasc} & MCQ&17M\\
& InsuranceQA \cite{feng2015applying} & InsuranceQA& 24K\\
\midrule

Tweet & Signal-1M(RT) \cite{suarez2018data} & Twitter QA & 2.86M\\
\midrule

News & TREC-NEWS	 \cite{soboroff2018trec} & Article & 595K \\
Article & TREC Robust \cite{voorhees2005trec} &Article&69.9\\
& MLSUM \cite{scialom2020mlsum} & News Articles & 1.5M\\

\midrule
Argument & Arg-Microtexts Synthesis \cite{wachsmuth2018retrieval} & Argument & 8.67K	\\
Retrieval & Conversational Argument \cite{ajjour2020overview} & Conv. Argument & 382K\\
\midrule
Community QA& CQADupstack \cite{hofstatter2020improving} & Community QA & 457K\\

\midrule
Entity & DBPedia \cite{hasibi2017dbpedia} & Entity & 4.63M\\
\midrule
 & Fever \cite{thorne2018fever} &  &5.42M \\
Fact&Climate-FEVER \cite{diggelmann2020climate}	& &5.42M\\
Checking& SciFact \cite{wadden2020fact} & claims climate-change&5K	\\
\midrule
Search & MSLR-WEB10K \cite{qin2013introducing} & Search Queries& 10K\\
Query & MSLR-WEB30K \cite{qin2013introducing} & Search Queries& 30K\\
\midrule
\end{tabular}

\end{table}
\label{Sec:dataset}
\section{Current Challenges and Further Directions}
In this survey paper, we have provided an overview of the current state of the field of information retrieval, including conventional term-based retrieval, pioneering methods for semantic retrieval, query and document augmentation, lexical dependency model, topic model, multilingual retrieval model, and deep learning methods for semantic retrieval. We have also discussed the two stages of the retrieval process: retrieval and ranking.

Recent studies have delved into the utilization of pre-training objectives that are specifically tailored to the task of information retrieval (IR). For instance,\cite{lee2019latent} proposed the use of a large-scale document collection and the Inverse Cloze Task (ICT) for retrieval tasks, as a method to improve performance. \cite{guu2020retrieval} on the other hand, proposed a method to capture inner-page and inter-page semantic relations through the usage of Body First Selection (BFS) and Wiki Link Prediction (WLP) for passage retrieval within question-answering (QA) tasks. Additionally, \cite{ma2021prop} and \cite{ma2021pre} proposed the Representative Words Prediction (ROP) objective for pre-training, which has shown to yield significant improvements. However, it is still in the early stages of research to effectively design pre-training objectives that are highly suited to the task of IR, thus providing ample room for further exploration and experimentation.

 Recent studies have highlighted some of the main challenges and future directions for semantic retrieval that can be considered as a general view on the field.

Handling long-tail queries: A recent study has highlighted that handling long-tail queries, which are queries that are infrequent or rare, is a major challenge for semantic retrieval systems.

Leveraging pre-trained models: Pre-trained models, such as BERT \cite{devlin2018bert}, have been shown to be effective in many natural language processing tasks and are being increasingly used in semantic retrieval.

Handling multilingual retrieval: With the increasing amount of multilingual information available on the web, handling multilingual retrieval has become an important challenge for semantic retrieval 

As for the future directions, some recent papers have proposed:
Developing more sophisticated models for document representation, such as deep neural networks and transformer-based models to improve the effectiveness of semantic retrieval \cite{Huang2021Mixed}.

\section{Conclusions}
\label{sec:conc}
This survey provided a comprehensive overview of the state-of-the-art for semantic retrieval models, in the context of information retrieval. We covered a wide range of topics, from the early semantic retrieval methods, to the most recent neural semantic retrieval methods, discussing the connections between them. In terms of structure, our focus was on the key IR topics: first-stage and second-stage retrieval, neural semantic retrieval model learning. Additionally, the survey highlights the major difficulties and challenges in the field, and points for promising directions for future research. Overall,  this survey is expected to be useful for researchers interested in this challenging topic, providing inspiration for new ideas and further developments.

\vspace{6pt}

\authorcontributions{Conceptualization, K.H. and H.P.; Writing – original draft, K.H.; Writing – review \& editing, K.H and H.P.}

\funding{
This work is funded by FCT/MCTES through national funds and co-funded by EU funds under the project UIDB/50008/2020. 
}

\institutionalreview{\hl{Not applicable} 
}

\informedconsent{\hl{Not applicable.} 
}

\dataavailability{\hl{Data is contained within the article. } 
} 

\conflictsofinterest{\hl{The authors declare no conflict of interest.} 
} 
\end{paracol}
\reftitle{References}



\bibliography{jalal}

\end{document}